# A simple and robust method for characterization of afterpulsing in single photon detectors

Gerhard Humer, Momtchil Peev, Christoph Schaeff, Sven Ramelow, Mario Stipčević, Rupert Ursin

*Abstract*— Single photon detectors are important for a wide range of applications each with their own specific requirements, which makes necessary the precise characterization of detectors. Here, we present a simple and accurate methodology of characterizing dark count rate, detection efficiency, and afterpulsing in single photon detectors purely based on their counting statistics. We demonstrate our new method on a custom-made, free-running single photon detector based on an InGaAs based avalanche photo diode (APD), though the methodology presented here is applicable for any type of single photon detector.

## I. INTRODUCTION

Single photon detection at telecom wavelengths has attracted significant research efforts due to its numerous applications in metrology and telecommunications as well as in quantum optics where it is particularly relevant for Quantum Key Distribution (QKD).

Characterization of single photon detectors has become an important task in order to compare and select the right parameters for a specific application. Here we present a novel method for afterpulsing characterization, which uses a discrete, binned probability density function of the timing distances between the measured events. Based on the theoretical probability density function of time measurement events, as recorded by a perfect detector, which detects photons, generated by a light source at random times and independently one from the other, this method allows separating the imperfection in a very simple way. It even lets detector characterization using only the intrinsic dark counts. This method is a generalization of a procedure proposed in [1,2], which is specifically designed for characterizing detectors operating in gated mode with the objective to obtain a robust estimate of the various performance parameters, especially the afterpulsing probability. The advancement presented in this work extends the applicability to the free-running detection mode and allows using any light generation process if it can be approximated by a Poisson one. Importantly, this includes intrinsic dark counts of the detector or background counts from residual stray light. Our method only requires the time-binned statistical measurement of detection events and is easily realizable in hardware allowing for a quick assessment of single photon counting detectors. Fundamentally, similar to [2] it is based on a linear regression fit of the detection events' histogram in contrast to an approximation (second order Taylor series expansion) of the afterpulsing waiting probability suggested in [1]. Simultaneously in contrast to [2] a precise mathematical derivation of the waiting probability of detection events is put forward and the waiting probabilities characterizing the different classes of events (source photons, dark counts, afterpulsing) are systematically studied.

We have tested our results using a self-designed and implemented single photon detector featuring custom-made electronics together with a commercial Indium Gallium Arsenide/Indium Phosphide (InGaAs/InP) based single photon avalanche diode (SPAD), model PGA-400 (Princeton Lightwave Inc.). It is fiber-pigtailed and packaged in a 14 pin butterfly housing and is sensitive in the 0.95-1.65 µm wavelength range.

Our paper is structured as follows: We first present the principle experimental setup and the theoretical background of our method, followed by an illustration based on measurement results and discussion. Our analysis includes the afterpulsing probability as well as the dark count rate. The theoretical analysis is founded on the approach of [1]. The latter is however augmented by two Appendices (I and II), which clarify the meaning of the framework, although some may exist is other previous statistics related work. (These Appendices are included mainly to make the text self-contained.) The main novel theoretical derivations are presented in Appendix III and not given in the main text of the article to allow separation of methodological approach and application relevant material. As an illustration of the method we present in an Appendix IV the measured dark count rates and afterpulsing probabilities as a function of temperature and efficiency for the mentioned detector.

Finally it should be stressed that the article presents a probability framework that can applied to estimate model parameters. To make the paper logically closed we have intentionally left out all statistic considerations on sample sizes and respective confidence intervals of the model estimates. Clearly the latter are indispensable in any practical application of the suggested framework.

G. Humer and M. Peev are with Austrian Institute of Technology (AIT), Donau-City-Strasse 1, 1220 Vienna, Austria

C. Schaeff, S. Ramelow and R. Ursin are with Institute for Quantum Optics and Quantum Information (IQOQI), Austrian Academy of Sciences, Boltzmanngasse 3, A-1090 Vienna, Austria.

M.Stipčević is with Ruđer Bošković Institute, Bijenička 54, 10000 Zagreb, Croatia.

## II. Experimental setup

The general scheme for characterizing single photon detectors is shown in Fig. 1. We use a tunable CW laser source (VIDIA-DISCRETE) at -20 dBm and at a wavelength of 1550 nm, augmented with attenuators and power splitters to reach low enough light levels. The attenuation and laser power are measured with a power meter (RIFCOS 575L). The input is adjusted to about 1.7 M photons/s (calculated from source power and attenuation) and allows us additionally to roughly estimate the photon detection efficiency of the detector.

The source is connected to the detector via a single mode fiber SMF-28 (Corning) with a core diameter of 8.2 µm. The detection efficiencies have been measured with this fiber.

The output pulses from the photon detector are precisely measured with the time-to-digital converter (AIT TTM8000 ). This time-tagging-module (TTM) provides 8 independent input channels for continuous time of arrival measurements. In the basic mode, sufficient for our measurements, the timing resolution is 82 ps simultaneously on 8 channels. It can continuously deliver up to 25 MEvents/s to a computer. In high-resolution modes a resolution of less than 10 ps can be achieved simultaneously on 2 channels and down to 1 ps if one channel is used exclusively for Start and the other exclusively for Stop signals.

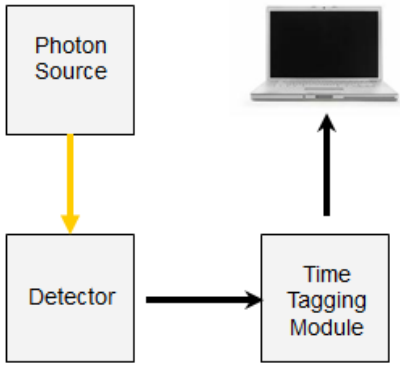

Figure 1. Sketch of the setup used in the measurements. The Photon Source can be switched OFF and ON to illuminate the Detector with light. All arrival times are recorded by the Time-Tagging Module and stored on a Computer.

The single photon detector implements a fast active quenching circuit in order to reduce the total avalanche charge and the negative effects due to slow restoring of the bias voltage after an avalanche. Moreover, an efficient thermoelectric cooling setup provides a stable temperature of below -60°C, which is necessary to achieve low dark count rates. By means of electrical adjustments of the quenching circuit, the module can be adjusted to operate with detection efficiency probabilities between around 0.3% and 10%. The dead-time is adjustable from 0.1 to 10 µs allowing for desired optimization of the trade-off between high peak count rates and low afterpulsing. Further the typical timing jitter is below 350 ps (FWHM) at about 10% efficiency measured at the standard SMA output connector.

## III. Method for statistical characterization of photon detectors

Originally [1, 2], time discretization has been considered, whereby the equidistant "time bins" have been defined as multiples of the gating period of the detector. Our first observation is that the concept of a time bin is well defined, whenever the number of time intervals elapsing after some event before the occurrence of a second one can be counted with a sufficient precision. This is also the case for a free running detector, if the elapsed time between a detection event and a subsequent one is measured using a time-tagging device, as shown in Fig. 2a.

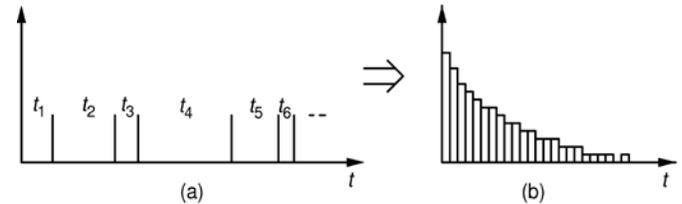

Fig. 2. Principle of the acquisition of time intervals. A sequence of time intervals $t_i$ as measured by the time tagging unit (a) is graphically illustrated as a histogram with finite bin width (b).

By means of the time-tagging unit the statistical distribution of waiting-times between two consecutive detection events can be precisely recorded. The recorded times can be graphically illustrated in a histogram as shown in Fig. 2b. This histogram represents a discrete approximation of the waiting-time interval distribution. The bin width of the histogram can, in principle, be chosen arbitrary, but there is a tradeoff between measurement time and approximation accuracy. With a bin size of 100ns and a measurement time of 10 minutes the histogram curve is already smooth enough to get good approximation results.

For uncorrelated events, the probability of a single detection event occurring $n$ time slots (time bins) after a preceding detection event can be expressed through the probabilities of such events occurring in single time slots. This holds true if the probabilities for detecting events in different time slots are independent from each other in time. This assumption is correct for an APD photon detector connected to a Poisson photon generation process(es) via a memory-less channel between them, as is the case in our setup. Furthermore we explicitly assume that the probability for detecting an event after the initial one is independent of the pre-history, i.e. that the detector state after registering a pulse is always the same. Clearly this assumption is not universally true and below we shortly necessary conditions for its validity, i.e. that the state of the detector after measuring a pulse is always the same.

The principle of our approach can be illustrated as follows. The probability that a detection event in a given time slot is followed by detection in some subsequent time slot, e.g. in the 3rd time slot, after the first one (the first event initiating the counting procedure and corresponding thus to time slot 0) can be expressed as:

$$P_H(3) = P(3)[1 - P(1)][1 - P(2)] \quad (1)$$

Here the probability of measuring the first subsequent event in the 3rd time slot, $P_H(3)$, is a product of the probabilities of no detection event in the first and second time slots and that of a detection event in the third time slot. Generally, the probability that the first subsequent event measured in time slot number $n$, is given by:

$$P_H(n) = [1 - P_{ne}(n)] \prod_{i=1}^{n-1} P_{ne}(i) \quad (2)$$

where the following notation has been used:
$P_H(n)$ probability of an event to occur $n$ time slots after a triggering one, with no detection events in between,
$n, i$ time slot indices,
$P_{ne}(n)$ probability of no detection event in the $n$-th time slot,
$P(n)$ probability of a detection event in the $n$-th time slot.
Note: $P(n) = 1 - P_{ne}(n)$.

With detection events due to source photons, dark counts and afterpulsing we get:

$$P_H(n) = [1 - (1 - P_S)(1 - P_d)(1 - P_a(n))] \times \prod_{i=1}^{n-1}[(1 - P_S)(1 - P_d)(1 - P_a(i))] \quad (3)$$

where:
$P_S$ probability to detect a source photon in one time slot,
$P_d$ probability to detect a dark count in one time slot,
$P_a(i)$ probability to detect an afterpulse count in the $i$-th time slot.

We note that $P_H(n)$ is a mass function of a discrete probability distribution defined over the positive integers $n = 1, 2, 3, ...$. (the proof is given in Appendix I). We further assume that source-photon detection events and dark count events can be described by a Poisson process with events occurring continuously and independently at a constant average rate:

$$P_H(n) = [1 - e^{-\mu_S} * e^{-\mu_d} * (1 - P_a(n))] \times \prod_{i=1}^{n-1}[e^{-\mu_S} * e^{-\mu_d} * (1 - P_a(i))] \quad (4)$$

or

$$P_H(n) = [1 - e^{-(\mu_S + \mu_d)}(1 - P_a(n))] \times e^{-(\mu_S + \mu_d)(n-1)} \prod_{i=1}^{n-1}[(1 - P_a(i))] \quad (5)$$

where:
$\mu_S = \eta \lambda_{S_0} \Delta t$ – the average number of detected source photons in the time window with $\eta \lambda_{S_0}$ being the rate of detected source photons, i.e. $\lambda_{S_0}$ – the rate of the source photons and $\eta$ – the detection efficiency, including any further attenuation, and $\Delta t$ the duration of the time slot;
$\mu_d = \lambda_d \Delta t$ – the average number of dark counts in the time window with $\lambda_d$ being the dark count rate and $\Delta t$ the duration of the time slot.

Here, similar to [1], we have taken into account that the distribution of events generated by a Poissonian process in any time window of duration $\Delta t$ is the Poisson distribution with mean equal to the average number of events in this window (e.g. $\mu_S = \eta \lambda_{S_0} \Delta t$ and $\mu_d = \lambda_d \Delta t$, with $\mu_S$ and $\mu_d$ being respectively the average number of detected source photons and dark counts in this time window). The probability of detection no photons from one of these sources in a $\Delta t$ time window is then equal to the 0-th term of the respective Poisson distribution, i.e. $e^{-\mu_S}$ or $e^{-\mu_d}$. Taking the logarithm of (5), we get:

$$\ln(P_H(n)) = \ln[1 - e^{-(\mu_S + \mu_d)}(1 - P_a(n))] - (\mu_S + \mu_d)(n - 1) + \ln\left\{\prod_{i=1}^{n-1}[(1 - P_a(i))]\right\}. \quad (6)$$

To demonstrate the application of (6) we consider two specific cases: detection with and without afterpulsing.

### A. Detection without afterpulsing ($P_a(n) = 0$)

Although this case is physically unrealistic it is instructive and will be used subsequently taking appropriate limits. For this case we get,

$$\ln(P_H(n)) = \ln(1 - e^{-(\mu_S + \mu_d)}) - (\mu_S + \mu_d)(n - 1) \quad (7)$$

or

$$\ln(P_H(n)) = -(\mu_S + \mu_d)n + \ln(1 - e^{-(\mu_S + \mu_d)}) + (\mu_S + \mu_d). \quad (8)$$

Clearly this is a linear function in $n$, $f(n) = \mu n + c$, where

$$\mu = -(\mu_S + \mu_d), \quad (9)$$

and

$$c = \ln(1 - e^{-(\mu_S + \mu_d)}) + (\mu_S + \mu_d). \quad (10)$$

The measurement procedure for this case is then as follows.
1. Switch OFF the photon source and collect sufficient data (due to dark counts) to obtain a statistically significant histogram. Then apply (8) to obtain $\mu$ using a linear regression. Since the source is switched off, $\mu_S = 0$ and one can easily calculate $\mu_d = \mu$.
2. Switch ON the Poisson photon source. Then apply (8-10) to determine $\mu = \mu_S + \mu_d$ using linear regression. Since $\mu_d$ has already been estimated in the previous step, we can then calculate $\mu_S = \mu - \mu_d$.

If $\lambda_{S_0}$, the rate of photons generated by the source, is independently measured, one can further obtain an estimate of the detection efficiency $\eta$ as:

$$\eta = \frac{\mu_S}{\Delta t \lambda_{S_0}} \quad (11)$$

We stress again that this simple characterization procedure is valid under the assumption that there is no afterpulsing, which is unphysical, but it still yields good approximate values in case of small or negligible afterpulsing probability.

### B. Detection with afterpulsing ($P_a(n) > 0$)

#### 1) $P_a(n)$ modeled with an exponential decay

A simple and realistic model of after pulsing [1] represents the probability density function $P_a(t)$ in (6) as a decreasing exponential of the elapsed time:

$$P_a(n\Delta t) = P_{a_0} e^{-\frac{n\Delta t}{\tau_0}}. \quad (12)$$

More elaborate studies [8] have shown that the decay can even more precisely be described by means of a sum of exponentials or a power function with a rational negative exponent. In any case all descriptions rely on a function that quickly decays with elapsed time. Equation (12) in particular assumes an exponential decay for the trapped carriers with effective de-trapping lifetime $\tau_0$ and associated amplitude $P_{a_0}$ which is related to the number of trapped carriers. Here, as above, $\Delta t$ is the bin width of the histogram, which as explained, is preferably taken to be equal to the time-tagging device time bin width. We mention that

$$P_a = \sum_{i=1}^{\infty} P_{a_0} e^{-\frac{i\Delta t}{\tau_0}} < 1, \qquad (13)$$

is the total probability for an afterpulse after detecting an event. The complementary probability $P_{na} = 1 - P_a$ is the probability of no afterpulse after a detection. Detector design naturally aims at low total after pulse probability. One technical means to do so is blocking the detector electrically after it fires when registering an event for a dead time $\tau_\delta = n_\delta \Delta t$, where we have assumed for convenience that the dead time is proportional to an integer number of time bins. Indeed, with dead time,

$$P_{a,\delta} = \sum_{i=n_d}^{\infty} P_{a_0} e^{-\frac{i\Delta t}{\tau_0}} = \sum_{i=1}^{\infty} P_{a_0} e^{-\frac{n_\delta \Delta t}{\tau_0}} e^{-\frac{i\Delta t}{\tau_0}} < P_a \qquad (14)$$

Note further that (12) is explicitly independent of the pre-history of the detector before firing the initiating event. Physically this means that any detector triggering yields always the same occupation of the trapped carrier energy levels. This verification of this assumption needs detailed investigation in order to check whether higher order effects might be relevant. I any case it is certainly satisfied if we have chosen a sufficiently high $\tau_\delta$, as otherwise it might be the case that a detector firing, soon enough after the first one may lead to even more dense occupation of trapped carrier levels. In what follows we explicitly assume that either the dead time is sufficiently long or that higher order effects are irrelevant and thus the basic assumption on the independence of our fundamental distribution $P_H(n)$ of pre-history holds, something that is typical of normal detector operation..

As shown in Appendix II when afterpulsing and dead time are considered, the discrete probability distribution $P_H(n)$ of registering a first event in the time slot $n$ after an initialization one the time slot 0 is to be replaced by discrete $P_{H,\delta}(n)$, $n = 1, 2, 3, ...$, which depends on the dead time and for which counting starts after the elapse of the dead time. It can be written (see (A.3-A.4)) as,

$$P_{H,\delta}(n) = \left[1 - e^{-(\mu_S+\mu_d)}\left(1 - P_{a_0}(\tau_\delta) e^{-\frac{i\Delta t}{\tau_0}}\right)\right] \times$$
$$e^{-(\mu_S+\mu_d)(n-1)} \prod_{i=1}^{n-1}\left[\left(1 - P_{a_0}(\tau_\delta) e^{-\frac{i\Delta t}{\tau_0}}\right)\right]. \qquad (15)$$

Here the afterpulsing probabilities even for low values of $n$ tend to 0 with the increase of the dead time and the description correspondingly tends to an afterpulsing free one, as can be expected intuitively. Correspondingly, in the logarithmic form (15) can be cast as

$$\ln\left(P_{H,\delta}(n)\right) = \ln\left[1 - e^{-(\mu_S+\mu_d)}\left(1 - P_{a_0}(\tau_\delta) e^{-\frac{n\Delta t}{\tau_0}}\right)\right]$$
$$- (\mu_S + \mu_d)(n-1) + R_\delta(n), \qquad (16)$$

where:

$$R_\delta(n) = \sum_{i=1}^{n-1} \ln\left(1 - P_{a_0}(\tau_d) e^{-\frac{i\Delta t}{\tau_0}}\right). \qquad (17)$$

With a direct numeric fit of both histograms (obtained with the source switched ON and OFF) one can in principle evaluate $\mu_S, \mu_d, P_{a_0}(\tau_\delta)$ and $\tau_0$. In this process it will be a significant advantage if one is able to reduce the potential ambiguity in numeric fitting an analytic expression of (20) and particularly a functional expression of the term $R_\delta(n)$. In [1] an approximation of this term to the second order has been obtained, but here we present instead a different approximation that is both intuitive and simple to apply. The basis of this approximation is the fact that for sufficiently large values of $i$ the corresponding terms in the sum in (17) quickly tend to zero and one can use a Cauchy convergence test to show that the sum itself approaches a constant, that is:

$$\lim_{n \to \infty} R_\delta(n) = R_{0,\delta} \qquad (18)$$

Thus, for sufficiently large $n$, the following approximation holds true:

$$\ln\left(P_{H,\delta}(n)\right) \approx -\mu n + \ln(1 - e^{-\mu}) + \mu + R_{0,\delta} \qquad (19)$$

where we have again denoted $\mu = \mu_S + \mu_d$ and taken into account that for the considered values of $n$, $1 - P_{a_0}(\tau_d) e^{-\frac{n\Delta t}{\tau_0}} \approx 1$. This is a linear function similar to that given in (8), whereby importantly the slope is given again by and the additive constant is now:

$$c_\delta = \ln(1 - e^{-\mu}) + \mu + R_{0,\delta}. \qquad (20)$$

Geometrically the graph of the function in (16) asymptotically tends to the linear function in (19). The important condition for the linearization to hold is that elapsed time $(n + n_\delta)\Delta t/\tau_0$ is sufficiently larger than the afterpulsing lifetime $\tau_0$, namely that afterpulses have virtually all died-off by the $n$-th time slot. For InGaAs/InP operating at temperatures higher than -50 °C, one can safely assume that virtually all afterpulses die off after ~5 µs [5-7]. In what follows we define this period to be a "maximum life time" $\tau$, after which e.g. there remains less than 5% of probability of afterpulsing events. This implies that $\tau \approx 3\tau_0$, which in turn gives $\tau_0 \approx 1.66$ µs for the discussed case.

A procedure to determine an estimate for the parameters $\mu$ and $c_\delta$ under the assumptions given above can be then summarized as follows:

1. Collect sufficient data to obtain a statistically significant histogram by measuring "in the dark", or, if the dark count rate is less than ~10-20 kcps, using a low-level light from a CW source so that the mean detection frequency is in the order of 10-20 kcps. (A discussion on the choice of this rate is given below.) Generate a histogram $P_H(n)$ with time intervals for up to 20 µs integrating for up to $10^6$ intervals to get sufficiently good statistics. *Note:*

*In case the dark counts rate is low, recording the histogram using low-level light contributes to speeding* up data acquisition.

2. Fit the linear approximation given in (16) for the chunk of the histogram $P_H(n)$ in 5-10 µs interval region. This will yield estimates of the constant parameters $\mu$ and $c_\delta$ in the region essentially free of afterpulses.
3. For the range of low to medium values of *n* (corresponding to time intervals between 0 and 5 µs) determine $P_{a_0}(\tau_\delta)$ and $\tau_0$ by using the explicit expressions in (16-17). Alternatively one can perform direct numeric fitting of the full curve. In any case, only two parameters ($P_{a_0}(\tau_\delta)$ and $\tau_0$) remain to be determined instead of four, a fact which greatly simplifies the task.

It is important to note that although the results, on which this procedure is based are exact, an experimental estimation of the parameters $\mu$ and $c_\delta$ is more precise for lower values of the Poissonian rate $\mu$. Indeed for a fixed overall histogram-recording time with the increase of $\mu$ the number of events that fall in the leftmost part of the histogram increases. Correspondingly, in this case the number of events that determine the linear part in a semi-log plot decreases leading to the increase of the fluctuation weight in this region. Hence, the parameters of the linear "tail" can be less stably determined. For this reason exact determination of the parameters is much better feasible for lower values of $\mu$. This is a clear indication that a low Poissonian event-generation frequency is required for applying (19-20) to get subsequently correct parameter estimation. We found that, as a rule of thumb, the average period between Poissonian events (dark counts + source photons) should be at least 10 times longer than the maximum lifetime $\tau$ of afterpulses. Taking the latter to be ~5 µs, as mentioned above, it is sufficient that the overall detection frequency be in the order of 10-20 kHz.

*2) Arbitrary or unknown model of the afterpulsing process*
Generally, afterpulsing can be more complex than in the simplified exponential model set forth in (12), [1]. For solid state avalanche photodiodes there is a convincing theoretical and experimental evidence that afterpulses are caused by one or more types of trapping centers each with its own trapping probability and lifetime [5]. In cases when one type of trap is predominant (as in [6]) our simple model may be sufficiently accurate.

A careful consideration of the method described above immediately reveals, however, that the explicit functional dependence of the afterpulsing time dependence is used only in very few steps of the parameter estimation procedure. The important aspect is that afterpulsing essentially fades out after a (relatively small) number of time slots and therefore the term $R_\delta(n)$ in (16) can be approximated by a constant after sufficiently many time sots allowing for the linearization of the equation, yielding (19-20). This holds true because from a physical point of view, irrespectively of the concrete model employed, afterpulsing is caused by trapped carriers that are in metastable states and these inevitably decay after a while. For this reason it is evident that (19) holds universally and can be used as demonstrated to determine the constant parameters $\lambda_s = \frac{\mu_s}{\Delta t}$, $\lambda_d = \frac{\mu_d}{\Delta t}$ and $c_\delta$.

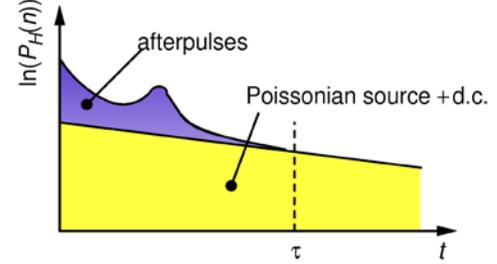

Figure 3. Example of a histogram (drawn in log scale) representing an arbitrary afterpulsing model whose important property is that afterpulses eventually die off after a time $\tau$. The histogram has a range due exclusively to dark counts and the Poissonian light source for $t > \tau$ and a range due to afterpulses for $t < \tau$.

An important parameter which can be determined robustly in this case is the time $\tau$, after which the experimental curve and the linear fit can no longer be differentiated, that's the time for which afterpulsing can be considered as already effectively "extinguished" (see Fig. 3). The corresponding time interval (or part of it) can then be used as e.g. a dead-time for applications that are sensitive to effects of afterpulsing (for example Quantum Key Distribution).

Regardless of whether the afterpulsing model is known or not, having determined the estimates $\mu$ and $c_\delta$ by linearly fitting (19-20) in the afterpulsing free region $t > \tau$ in Fig. 3 one can directly get a lower bound of the total afterpulsing probability. Referring to the analysis of the total (cumulative) probability of afterpulsing in general, presented in APPENDIX III (cf. (A 3.11)), it is then straightforward to see that

$$P_a < 1 - e^{R_{0,\delta}} = 1 - e^{(-\ln(1-e^{-\mu})-\mu+c_\delta)}, \qquad (21)$$

where in the last step we have taken into account (20), which (as stated) holds independently of the afterpulsing mechanism, provided that the latter is compatible with the general assumptions discussed above.

Respectively, the number of afterpulses per Poisonnian photon (source photon or dark count) can bounded as follows (see APPENDIX III),

$$N_{\frac{a}{s,d}} \leq \frac{\overline{P}_a}{\underline{P}_{s,d}} < \frac{1-\underline{P}_{s,d}}{\underline{P}_{s,d}} = \frac{1-e^{(-\ln(1-e^{-\mu})-\mu+c_\delta)}}{e^{(-\ln(1-e^{-\mu})-\mu+c_\delta)}} =$$
$$= e^{(\ln(1-e^{-\mu})+\mu-c_\delta)} - 1. \qquad (22)$$

In this case, following the approach detailed in Appendix III, we can also obtain more detailed information about afterpulsing, namely the waiting probability of afterpulsing. We have demonstrated (see (A3.7)) that

$$P_{H;a}(n) < P_H(n) - e^{-n\,\mu + c_{0,\delta}}, \qquad (23)$$

where we have again used (20) on the same grounds as above. It must be stressed, however, that a segment-wise lower bound of afterpulsing probability density function can also be obtained using recursive relations that generally follow from an approach analogous to the derivation of (3) but lie outside the scope of the present paper. We note in passing that the possibility of such an approach has been mentioned and initial calculations have been carried out in [2]. Unfortunately the model the authors use is only approximate in terms of per-slot event probability (cf. (2) of [2] and compare to (5) in this article) for which reason the results in [2] on the afterpulsing pdf are inaccurate.

A procedure for characterization of afterpulses on case of general or unknown afterpulsing model is as follows:

1. Record time intervals using dark counts only or, if dark count rate is less than ~10-20 kcps add a small intensity of light from the CW source so that the mean detection frequency is on the order of 10-20 kcps. Fill the histogram $P_H(n)$ with time intervals up to 20 µs integrating for up to $10^6$ intervals to achieve sufficiently good statistics.
2. Fit the linear approximation given in (16) to the part of the histogram $P_H(n)$ in the region [$\tau$, 20 µs], in our case $\tau$ = 5 µs. The fit will yield estimates on parameters $\mu$ and $c_{0,\delta}$, assuming that the fit region is virtually afterpulse-free.
3. Use (21) to determine a lower bound of the total afterpulsing probability $P_a$ and (22) to determine the number of afterpulses per "trigger" (Poissonian) pulse. NOTE: These results are general and refer to both cases of known and unknown afterpulsing mechanisms / models
4. One may further optimize $\tau$ and start with a lower value (e.g. $\tau$ = 0.5 µs) and evaluate the upper bound of $P_a$ as a function of $\tau$ for a series of equidistant values (e.g. 1 µs, 1.5 µs, 2 µs, 2.5 µs, ..., 10 µs). As $\tau$ rises, also the estimated bound of $P_a$ changes eventually approaching a constant value, which is exactly the optimal estimate of the bound.
5. Determine a lower bound of the per time slot waiting probability of afterpulsing using (21) and (22). (As indicated above this might be used for deriving a upper bound of the afterpulsing probability density function.)

IV. PERFORMANCE TESTS OF A SINGLE PHOTON DETECTOR

We now use the setup described above (Fig. 1) as an experimental procedure for statistical characterization described above to test the performance of our custom-made single photon detector. We will use the method of detection with an unknown model of afterpulsing (Section III.B.2)) to demonstrate the most general procedure.

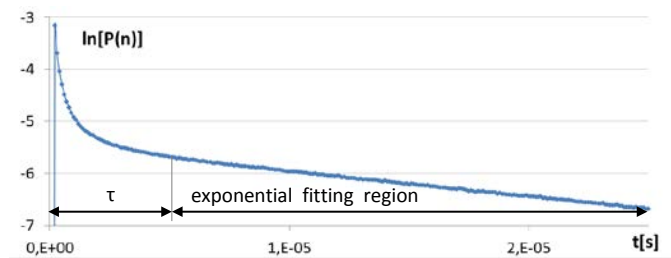

Figure 4. Measured histogram using the detector's dark counts. Two regions are depicted: one containing virtually all afterpulses (from 0 to $\tau$) and other containing virtually only either real photon detections or dark counts (from $\tau$ to 20 µs) are depicted.

Step 1. First, we measured time intervals between detector events induced by the detector dark counts. The dark count rate (including afterpulses) was 7390 cps allowing for rapid acquisition of $10^6$ intervals, using $\tau$ = 5 µs.

Step 2. The linear regression (fit) of (19) in the interval [$\tau$, 20 µs] yields: $\mu$ = 0,476/µs and $c_\delta$ = 5,49. (Here $\tau_\delta$ = 0.1 µs has been used.)

Step 3. In order to determine the total afterpulsing probability we have used (21) and obtained $P_a < 15.7\%$.

Step 4. By taking shorter and longer values in the range 4-8 µs for $\tau$ and repeating steps 2 and 3, we obtained mutually consistent values for the upper bound of $P_a$ within the experimental errors and concluded that the value of $\tau$ = 5 µs is acceptable.

Finally, as an illustration of usefulness of the described characterization procedure, we have optimized the duration of the dead time required to reduce the total afterpulsing probability to less than 1%. We found that elongating the dead time from present 0.1 µs to 3.0 µs would reduce the afterpulsing probability from 13.5% to 0.98%. The waiting probability in this case is shown in Fig. 5.

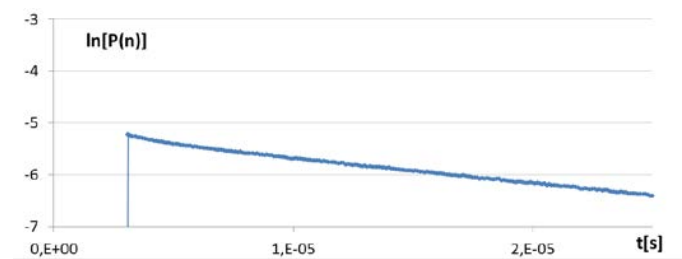

Figure 5. The histogram of the waiting probability with a dead time of 3.0 µs

V. CONCLUSION

In this paper, we have presented a new methodological (theoretic and experimental) framework to characterize the afterpulsing behavior in single photon detectors in free running mode, purely based on the counting statistics of these detectors. The methodology builds on existing work but is based on a precise mathematical formulation that was lacking in previous attempts (see Sections I and III for a comparison of our results with [1] and [2]). Bounds and estimate-accuracy are discussed in detail. We have presented some illustrations of our approach, particularly an upper bound of the afterpulsing probability, the estimate being reliable, and moreover easy to apply as no independent light source is required at all.

The methodology can be used in subsequent work in the field, for an in-depth analysis of arbitrary avalanche photodiodes in free-running mode by simple technical means. A particular example to this end is obtaining an estimate of the afterpulsing p.d.f. as briefly outlind in the text.

APPENDIX I

Here we prove that the function $P_H(n)$, defined in (3), the waiting probability distribution is indeed a mass function of a discrete probability distribution, defined over the integers $n = 1, 2, 3, \ldots$. All quantities $P_H(n)$ for all $n$ are strictly positive numbers that can be interpreted as the probability for a detector firing event *to occur for the first time* in time slot $n$. Formally one then needs only to prove that

$$\sum_{n=1}^{\infty} P_H(n) = 1. \tag{A1.1}$$

We shall do so in a somewhat indirect way, which however has the advantage of giving an alternative interpretation of the concept of *waiting probability*. Let us first consider a finite set of $N$ independent random experiments, each yielding success (1) with probability $p_n$, and failure (0) with probability $q_n$, with $p_n$ and $q_n$, depending on the number of the experiment $n = 1, 2, 3, \ldots, N$. Naturally for each $n$,
$$p_n + q_n = 1. \tag{A1.2}$$
The probability for a particular sequence of outcomes, e.g. $0_1 1_2 \ldots 1_N$, one of $2^N$ in number, is $q_1 p_2 \ldots p_N$,
$$P(0_1 1_2 \ldots 1_N) = q_1 p_2 \ldots p_N. \tag{A1.3}$$
The probability for an outcome, such as $0_1 1_2 \ldots 0_{k-1} 1_k X$, $k < N$, i.e. the set of $2^{N-k}$ strings with fixed first $k$ positions but arbitrary subsequent positions is

$$P(0_1 1_2 \ldots 0_{k-1} 1_k X) =$$
$$= q_1 p_2 \ldots q_{k-1} p_k (p_{k+1} + q_{k+1}) \ldots (p_N + q_N) =$$
$$= q_1 p_2 \ldots q_{k-1} p_k. \tag{A1.4}$$

$$0_1 1_2 \ldots 0_{k-1} 1_k X = \bigcup_{k+1}^{N} 0_1 1_2 \ldots 0_{k-1} 1_k s_i, s_i = \{0,1\}. \tag{A1.5}$$

It is now easy to see that the probability distribution we are looking for is the following,

$$P_H(n) = P(q_1 q_2 \ldots q_{n-1} p_n X) = p_n \prod_{i=1}^{n-1} q_i. \tag{A1.6}$$

In other words it the probability for finding the set of strings $0_1 0_2 \ldots 0_{n-1} 1_n X$, fr which the firs $n-1$ tests bring failure and success for the first time in the $n^{th}$ test. What happens after the $n^{th}$ test is immaterial as the detector resets a new series of the same time of experiment starts. It is thus clear that for any fixed. This is the "waiting probability" for any finite $N$.

Clearly all possible strings are covered by the sets $0_1 0_2 \ldots 0_{n-1} 1_n X$, for $n = 1, 2, 3, \ldots, N$ with one single exception. Clearly the string $0_1 0_2 \ldots 0_N$ of $N$ failures is not contained in it. It follows immediately that

$$\sum_{n=1}^{N} P_H(n) = 1 - P(0_1 0_2 \ldots 0_N) = 1 - q_1 q_2 \ldots q_N. \tag{A1.7}$$

The probability distribution with infinite but discrete domain $n = 1, 2, 3, \ldots$, can be seen as the limit of the finite discrete distribution $P_H(n)$, $n = 1, 2, 3, \ldots, N$, for $N \to \infty$. For the infinite distribution the following limit holds

$$\sum_{n=1}^{\infty} P_H(n) = \lim_{N \to \infty} \sum_{n=1}^{N} P_H(n) =$$
$$= 1 - \lim_{N \to \infty} q_1 q_2 \ldots q_N = 1. \tag{A1.8}$$

The relation $\lim_{N \to \infty} q_1 q_2 \ldots q_N = 0$, follows from the fact that this is a strictly positive monotonously decreasing sequence positive but smaller than one numbers, for which for any arbitrarily small positive number $a$ there exists an index $N_0$ such that $q_1 q_2 \ldots q_{N_0} < a$.

Relation (A8.1) is what was needed to be proven. Thus $P_H(n)$, $n = 1, 2, 3, \ldots$ is a probability mass function of an infinite discrete distribution. It can be interpreted as an infinite series of success-failure experiments or a "coin toss" series with different but fixed probabilities for success or failure in each coin toss. Each separate event in this distribution is the set all tosses that have their first success in the $n^{th}$ toss.

APPENDIX II

Substituting (12) in (5), without taking into account the dead time, we get:

$$P_H(n) = \left[1 - e^{-(\mu_S + \mu_d)} \left(1 - P_{a_0} e^{-\frac{n\Delta t}{\tau_0}}\right)\right] \times$$
$$e^{-(\mu_S + \mu_d)(n-1)} \prod_{i=1}^{n-1} \left[\left(1 - P_{a_0} e^{-\frac{i\Delta t}{\tau_0}}\right)\right] \tag{A2.1}$$

A description that also involves the dead time is fundamentally similar,

$$P_{H,\delta}(n) = \left[1 - e^{-(\mu_S + \mu_d)} \left(1 - P_{a_0} e^{-\frac{n\Delta t}{\tau_0}}\right)\right] \times$$
$$e^{-(\mu_S + \mu_d)(n - n_\delta - 1)} \prod_{i=n_\delta+1}^{n-1} \left[\left(1 - P_{a_0} e^{-\frac{i\Delta t}{\tau_0}}\right)\right];$$
$$\text{for } n > n_\delta,$$
$$P_{H,\delta}(n) = 0; \quad \text{for } n \leq n_\delta, \tag{A2.2}$$

as for all time slots before the dead time has elapsed the probability for detecting an event is physically fixed to be zero.

Obviously (A2.2) reduces to (A2.1) if counting starts with the first time slot after the dead time and if $P_{a_0}$ is replaced with $P_{a_0}(\tau_\delta) = P_{a_0} e^{-\frac{n_\delta \Delta t}{\tau_0}}$. Indeed the discrete probability distribution $P_{H,\delta}(n)$, $n = 1, 2, 3, \ldots$, for which counting starts after the elapse of the dead time can be written as,

$$P_{H,\delta}(n) = \left[1 - e^{-(\mu_S + \mu_d)} \left(1 - P_{a,\delta}(n)\right)\right] \times$$
$$e^{-(\mu_S + \mu_d)(n-1)} \prod_{i=1}^{n-1} \left[\left(1 - P_{a,\delta}(i)\right)\right], \tag{A2.3}$$

$$P_{a,\delta}(i) = P_{a_0}(\tau_\delta) e^{-\frac{i\Delta t}{\tau_0}}, \tag{A2.4}$$

(compare with (14)).

APPENDIX III

Here we present a short general analysis of probability of afterpulsing based on the assumption that afterpulsing probability decays sufficiently quickly after an initial excitating event.

To do this end we first consider the probabilities for different events in a single time slot. Obviously, the following single-slot events are feasible a-priori: i) detection of no-event; ii) arrival and detection of a Poissonian event alone and no afterpulse; iii) arrival and detection of an afterpulse and no Poisonnian event; iv) arrival of both a Poisoinian event and an afterpulse and detection of one of these (it is of course impossible to differentiate which one has been really detected). Clearly then, the measure of ii) can be seen as a lower bound for the probability to detect a Poissonian event in this time slot and the measure of ii) + iv) as an upper bound for the probability to detect a Poissonian event in this time slot.

As we have seen in Appexdix I, the waiting probability, i.e. probability to get the first counting event after n slots is a well-defined one. Following the pattern outlined for the single slot events we easily get the corresponding classification of first counting events in slot n. A priori, the first event can be $i)_H$ a Poisonnian event with no afterpulsing event, $ii)_H$ an afterpulsing event with no Poissonian event. However, there is also the probability of $iii)_H$ simultaneous arrival and resp. detection of both a Poissonian photon and an afterpulsing event in the same time slot (whereby naturally only one of these is taken into account in an experiment, but there is no meaningful of differentiation which one of these two).

Correspondingly, we can readily define lower and upper bounds for the waiting probability for the first detected event to be a Poisonian one, namely the measure of $i)_H$ is the lower bound, while the measure of $i)_H + iii)_H$ is the upper bound. We denote these lower and upper bounds with $\underline{P}_{H;s,d}(n)$ and $\overline{P}_{H;s,d}(n)$, respectively.

Following the arguments that lead to (5) it is straight forward to see that

$$\underline{P}_{H;s,d}(n) = (1 - P_a(n))(1 - (1 - P_s)(1 - P_d)) \times$$
$$e^{-(n-1)(\mu_S+\mu_d)} \prod_{i=1}^{n-1}\left[\left(1 - P_{a,\delta}(i)\right)\right], \quad (A3.1)$$

$$\overline{P}_{H;s,d}(n) = (1 - (1 - P_s)(1 - P_d)) \times$$
$$e^{-(n-1)(\mu_S+\mu_d)} \prod_{i=1}^{n-1}\left[\left(1 - P_{a,\delta}(i)\right)\right], \quad (A3.2)$$

or

$$\underline{P}_{H;s,d}(n) = \left(1 - P_{a,\delta}(n)\right)\overline{P}_{H;s,d}(n) \quad (A3.3)$$

where $P_{a,\delta}(n)$ is the dead-time dependent afterpulsing probability, which has a clear intuitive meaning and is a quantity that can be experimentally evaluated in principle even if the exact afterpulsing model is unknown. It is further clear that,

$$\underline{P}_{H;s,d}(n) = \left(1 - e^{-(\mu_S+\mu_d)}\right)e^{-(n-1)(\mu_S+\mu_d)} \times$$

$$\prod_{i=1}^{n}\left[\left(1 - P_{a,\delta}(i)\right)\right]; \quad (A3.4)$$

$$\overline{P}_{H;s,d}(n) > \underline{\underline{P}}_{H;s,d}(n) =$$
$$\left(1 - e^{-(\mu_S+\mu_d)}\right)e^{-(n-1)(\mu_S+\mu_d)+ R_{0,\delta}}, \quad (A3.5)$$

where $\underline{\underline{P}}_{H;s,d}(n)$ is a simple to calculate lower estimate of the lower bound $\underline{P}_{s,d}(n)$ and $R_{0,\delta}$ is defined as in (18).

Respectively, the probability of afterpulsing $P_{H;a}(n)$ to be the first registered event upper-bounds the probability measure of ii) – $\underline{P}_{H;a}(n)$, but lower-bounds the probability of ii) + iii) – $\overline{P}_{H;a}(n)$,

$$\underline{P}_{H;a}(n) \leq P_{H;a}(n) \leq \overline{P}_{H;a}(n),$$
$$\underline{P}_{H;a}(n) = P_H(n) - \overline{P}_{H;s,d}(n),$$
$$\overline{P}_{H;a}(n) = P_H(n) - \underline{P}_{H;s,d}(n) < \overline{\overline{P}}_{H;a}(n)$$
$$= P_H(n) - \underline{\underline{P}}_{H;s,d}(n). \quad (A3.6)$$

These inequalities, together with (A3.5) readily imply that,

$$P_{H;a}(n) < P_H(n) - \left(1 - e^{-(\mu_S+\mu_d)}\right)e^{-(n-1)(\mu_S+\mu_d)+ R_{0,\delta}}. \quad (A3.7)$$

It is of course important to know by how much $\overline{\overline{P}}_{H;a}(n)$ exceeds the upper bound $\overline{P}_{H;a}(n)$, i.e. what is the absolute error $E_{H;a}(n)$ per time slot is, in case $\overline{\overline{P}}_{H;a}(n)$ is used as an estimate of $P_{H;a}(n)$. From (A3.4-A3.7) then we readily get that

$$E_{H;a}(n) = \overline{\overline{P}}_{H;a}(n) - \overline{P}_{H;a},$$
$$E_{H;a}(n) = \left(1 - e^{-(\mu_S+\mu_d)}\right)e^{-(\mu_S+\mu_d)(n-1)} \times$$
$$\left\{\prod_{i=1}^{n-1}\left[\left(1 - P_{a,\delta}(i)\right)\right] - \prod_{i=1}^{n}\left[\left(1 - P_{a,\delta}(i)\right)\right]\right\}. \quad (A3.8)$$

Correspondingly the cumulative lower and upper probability bounds for the first detected event to be a Poissonian and not an aftepulsing one can be defined as $\underline{P}_{s,d}$ and $\overline{P}_{s,d}$ with

$$\underline{P}_{s,d} = \sum_{i=1}^{\infty}\left(1 - e^{-(\mu_S+\mu_d)}\right)e^{-(i-1)(\mu_S+\mu_d)} \times$$
$$\prod_{k=1}^{i}\left[\left(1 - P_{a,\delta}(k)\right)\right], \quad (A3.9)$$

$$\overline{P}_{s,d} = \sum_{i=1}^{\infty}\left(1 - e^{-(\mu_S+\mu_d)}\right)e^{-(i-1)(\mu_S+\mu_d)} \times$$
$$\prod_{k=1}^{i-1}\left[\left(1 - P_{a,\delta}(k)\right)\right]. \quad (A3.10)$$

Moreover, the following inequalities hold

$$\underline{\underline{P}}_{s,d} < \underline{P}_{s,d} \leq P_{s,d} \leq \overline{P}_{s,d}, \quad (A3.11)$$

with $P_{s,d}$, being the cumulative probability that the first detected event is a Poissonian one and

$$\underline{P}_{s,d} = \sum_{i=1}^{\infty}(1 - e^{-(\mu_S+\mu_d)})e^{-(i-1)(\mu_S+\mu_d)+R_{0,\delta}}$$

$$= \frac{(1 - e^{-(\mu_S+\mu_d)})e^{R_{0,\delta}}}{1 - e^{-(\mu_S+\mu_d)}} = e^{R_{0,\delta}}. \quad (A3.12)$$

Respectively, the cumulative probability of the first detected event to be an afterpulsing one $P_a$, together with the respective upper and lower bounds ($\overline{P}_a$ and $\underline{P}_a$), satisfies inequalities analogous to those given in (A3.6),

$$1 - \overline{\overline{P}}_{s,d} = \underline{P}_a \leq P_a \leq \overline{P}_a = 1 - \underline{P}_{s,d} < \overline{\overline{P}}_a$$
$$\overline{\overline{P}}_a = 1 - \underline{\underline{P}}_{s,d}. \quad (A3.13)$$

This equation, together with (A3.12) readily implies that,

$$P_a < \overline{\overline{P}}_a = 1 - e^{R_{0,\delta}}. \quad (A3.14)$$

The cumulative error $E_{H;a}$ of using the estimate $\overline{\overline{P}}_{H;a}$ instead of $\overline{P}_{H;a}$, follows from (A3.8),

$$E_a = \overline{\overline{P}}_a - \overline{P}_a = \sum_{i=1}^{\infty} E_{H;a}(i),$$

$$E_a = (1 - e^{-(\mu_S+\mu_d)}) \times$$
$$\sum_{i=1}^{\infty} e^{-(\mu_S+\mu_d)(i-1)} \left\{ \prod_{k=1}^{i-1}\left[(1 - P_{a,\delta}(k))\right] - \prod_{k=1}^{i}\left[(1 - P_{a,\delta}(k))\right] \right\}. \quad A3.15)$$

Unfortunately we could not find a way to present the right hand side in a closed analytical form. What is obvious, however, is that each term in the sum has a Poissonian dependent part and an afterpulsing dependent one. The Poissonian dependent terms form a series that would sum to 1 if the afterpulsing part would be equal to one. This a geometric series that for high Poissonian rate has higher values for the lower-index part of the series and lower values for the higher index part of the series, i.e. it converges to 1 quicker in comparison to the terms with the same index for lower values of the Possonian rate. The afterpulsing related series in curly brackets are decreasing extremely quickly (hyper-exponentially) to 0. In this sense the products of the afterpulsing terms with the Poisonian ones leads to a series for which the higher index terms are essentially cancelled out and the lower ones prevail. For this reason higher Poissonian rates lead to higher total error $E_a$. We have performed some numerical analysis with Poissonian rates ranging from 10kHz and 20 MHz and typical afterpulsing constants in a single exponential decay model and have got total error rates ranging from $5*10^{-7}$ to $3*10^{-5}$. The fact that the total error rate increases with the Poissonian rate is generally insignificant as the absolute value of the increase can safely be neglected. Note, however, that (as discussed in Section III.B.2)) that from an experimental point of view it is strictly a must to work with low Poissonian rates. For high rates, on a scale that cause Poissonian events *before* the afterpulsing from a previous Poisssonian event extinguishes, the experimental determination of parameters gets imprecise due to statistical fluctuations.

APPENDIX IV

Here we shortly present some results on the performance of the self-designed detector mentioned in the Introduction.

### A. Afterpulsing

We have measured the afterpulsing probability for different values of detector cooling temperature and the quantum efficiency. As well known the afterpulsing increases with quantum efficiencies and with the decrease of the cooling temperature (as a consequence that the decay time for trapped carriers increases when temperature falls).

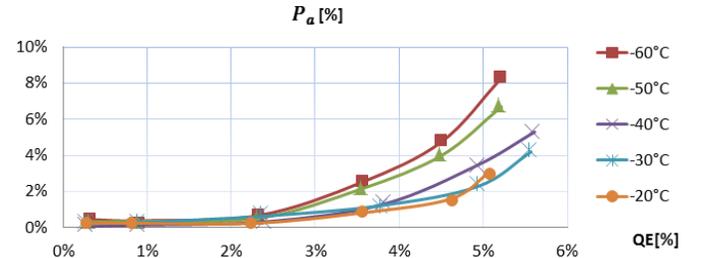

Figure 6. Estimated percentage of afterpulsing vs. efficiency at different temperatures (3µs deadtime)

### B. Dark count rate

Here we present the results for the variation of the dark count rate with detector cooling temperature and quantum efficiency. As it is also well known, dark count rate increases with both temperature and with quantum efficiency.

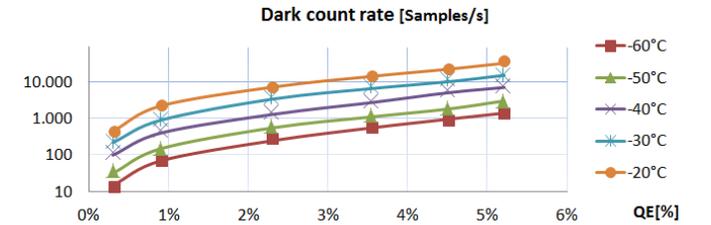

Figure 7. Dark count rate vs. efficiency measurement at different temperatures

ACKNOWLEDGMENT

We gratefully acknowledge grants from the European Space Agency (Contract 4000104180/11/NL/AF), the FFG for the QTS project (No. 828316) and European Commission grant Q-ESSENCE (No. 248095). MS is funded by MoSES 098-0352851-2873. SR is funded by an EU Marie-Curie Fellowship (PIOF-GA- 2012- 329851). CS acknowledges the doctoral program CoQuS (W1210-2).

One of the authors, MP, acknowledges very fruitful conversations with Thomas Lorünser.